\documentclass[12pt,preprint]{aastex}
\input{epsf}
\begin{document}
\def\msun{\hbox{${\cal{M}}_{\odot}$}}
\def\massA{\hbox{${\cal{M}}_A$}}
\def\massB{\hbox{${\cal{M}}_B$}}
\def\massAB{\hbox{${\cal{M}}_{A+B}$}}

\title{Binary Star Orbits. V. \\ The Nearby White Dwarf - Red Dwarf pair 40 Eri BC}

\author{Brian D.\ Mason, William I.\ Hartkopf and Korie N.\ Miles\altaffilmark{1}}
\affil{U.S. Naval Observatory \\
3450 Massachusetts Avenue, NW, Washington, DC, 20392-5420 \\
Electronic mail: (brian.d.mason, william.hartkopf)@navy.mil}


\altaffiltext{1}{SEAP Intern.}

\begin{abstract}

A new relative orbit solution with new dynamical masses is 
determined for the nearby white dwarf - red dwarf pair 40 Eri BC. 
The period is 230.09$\pm$0.68y. It is predicted to close slowly over
the next half-century getting as close as 1\farcs32 in early 2066. 
We determine masses of 0.575$\pm$0.018 \msun ~for the white dwarf 
and 0.2041$\pm$0.0064 \msun ~for the red dwarf companion. The 
inconsistency of the masses determined by gravitational redshift and
dynamical techniques, due to a premature orbit calculation, no 
longer exists.

\end{abstract}


\section{Introduction}

One of the more widely separated physical multiples in the sky, 40
Eri consists of a nearby, naked-eye star (HR 1325A) and 
a closer pair (BC) sharing the same, very large, proper motion over 
a minute of arc away. Parameters for the multiple system are 
presented in Table 1. In that table, Column 1 provides the relevant 
parameter, Columns 2, 3 and 4 gives the value for A, B and C, 
respectively, while Column 5 gives the reference(s). Note that we do
not give the position for C although Table 5 does provide the 
$\delta$ from the B position. This multiple system was listed as 
\#518 in F.G.W.\ Struve's (1837) catalog of double stars. Due to the
immensity of this catalog and it's logical structure the star number
in this catalog is taken as it's $``$discovery designation" despite 
being measured first by William Herschel (1785) almost 50 years 
earlier. The first accurate observation would wait another 14 years 
(Dawes 1867) after Struve's catalog. The AB pair, having only 
changed its position angle by 6$^{\circ}$ since its first measure 
233 years ago, would have a very long orbital period. However, BC 
was recognized as more rapidly moving and interesting. This interest
went beyond just being a potentially faster moving orbit pair when 
Adams (1914) noted it as $``$an A-type star of very low luminosity,"
i.e., a white dwarf. It appears and is described as an outlier in 
one of the very first color-luminosity diagrams (Russell 1914, see 
Figure 1). The star is, in fact, the second brightest known white 
dwarf, with an apparent magnitude V = 9.53 (Kidder et al.\ 1991); 
versus V = 8.44 for Sirius B (Bond et al.\ 2017). It is also by far 
the easiest to see, as Sirius B is lost in the glare of its primary 
(Bond et al.\ 2017), while the primary here is not only fainter (V =
4.43; Ducati 2002), but much farther from its companion 
($\rho~\sim~83\farcs7$). 

Due to the long period of most visual binaries (and the 
understandable impatience of calculators), orbits are often 
calculated when they $``$can" be and not necessarily when they 
$``$should" be. The first known orbit of the pair was by Gore 
(1886). In the Catalogue of Visual Binary Star Orbits (Finsen 1934),
the preferred orbit for 40 Eri BC was that of van den Bos (1926) as 
it was in the 2$^{\it nd}$ Catalogue (Finsen 1938). By the time of 
the 3$^{\it rd}$ Catalogue (Finsen \& Worley 1970), the preferred 
orbit was Orbit III of Wielen (1962), and this was updated again for
the 4$^{\it th}$ Catalog (Worley \& Heintz 1983), where the 
preferred orbit was that of Heintz (1974). It remained so in the 
5$^{\it th}$ Catalog (Hartkopf et al.\ 2001) and later electronic 
catalogs until the current calendar year. Heintz's (1974) mass 
estimates were 0.43$\pm$0.02 \msun ~for the white dwarf and 
0.16$\pm$0.01 \msun ~for the M dwarf companion. Using the modern 
Hipparcos parallax (van Leeuwen 2007) the masses would be 
0.48$\pm$0.02 \msun ~for the white dwarf and 0.17$\pm$0.01 \msun 
~for the M dwarf companion. 

Unfortunately, the dynamical mass of the white dwarf was rather 
different from the result obtained through analysis of the 
gravitational redshift, for example, 0.53$\pm$0.04 \msun ~from
Koester \& Weidemann (1991). Indeed, much ink has been spilled 
seeking to reconcile the differences between these two approaches
(Koester et al.\ 1979, Wegner 1979 \& 1980, Reid 1996, Provencal et 
al.\ 1998). 

\section{Measures of 40 Eri BC}

\subsection{New Measures}

The pair is suitable for observation by the USNO speckle camera on 
the 26$''$ refractor in Washington (Mason et al.\ 2011a,b) at the
suggestion of Howard Bond the pair was repeatedly observed until it
was too far off the meridian at twilight. Observed three times per 
night on six different nights, the calibration and methodology are 
as described in Mason \& Hartkopf (2017). The mean positions from 
these observation are presented in Table 2. In that table, Columns 
1, 2, 3, 4 and 5 provides the mean epoch of observation (in 
fractional Julian year), the position angle (in degrees), its error 
(in degrees), the separation (in seconds of arc), and its error (in 
seconds of arc). Note that the position angles have not been 
corrected for precession, and are thus based on the equinox for the 
epoch of observation. Column 6 gives the number of nights in the 
mean position and Columns 7 and 8 provide residuals to the orbit 
presented in \S 3. The $``$weight" of each measure used in the orbit
solution is given in Column 9 while Column 10 identifies the source 
of the observation.

The mean intranightly error is 0\fdg04 for the position angle 
($\theta$) and 0\farcs0039 for the separation ($\rho$). The errors
presented for position angle and separation presented in Table 2 are
the internightly errors\footnote{The error in position angle 
for the USNO speckle measures are not zero, but round to 0.0 when 
given at the precision of the measure.}.

The pair will be observable again in mid-September, but as described
in \S 3 below the accumulation of additional data will only 
make minute incremental improvement until, probably, the second half
of the 21$^{\it st}$ Century.

Also presented in Table 2 are measures obtained by matching the
components with objects in large catalogs with reliable astrometry. 
Using the same methodology as described in Wycoff et al.\ (2006) the
pair was matched with the 2MASS Point Source Catalog\footnote{2003 
all-sky release. See {\tt Vizier On-line Data Catalog: II/246}.}.
Similarly, the pair was matched with UCAC4 (Zacharias et al.\ 2013) 
using the techniques described in Hartkopf et al.\ (2013). Errors, 
when they can be determined from multiple measures, are presented as
well.

\subsection{Measures from the WDS}

Measures used in the orbit solution (\S3), from the Washington 
Double Star Catalog (hereafter, WDS, Mason et al.\ 2001) are 
presented in Table 3. In this table Columns 1, 2, and 3 provide the
mean epoch of observation, position angle and separation. Again, the
position angles are for the equinox of the epoch of observation. 
Column 4 lists the number of nights in the mean position, Columns 5 
\& 6, the O$-$C residuals to the orbit, while Column 7 is the 
$``$weight" used in the orbit solution. Column 8 is the source of 
the  measure and Column 9 is reserved for notes.

Despite IAU resolutions (IAU 1977) recommending that observations be
published using dates given in Julian epoch (JE), classic double 
star data have primarily been published with the date of observation
given at the fractional Besselian epoch (BE). We are in the process 
of evaluating the 9341 references used in the compilation of the WDS
and adjusting the observation epoch from BE to JE when appropriate. 
Accordingly, the measures listed in Table 3 have been converted to
Julian epoch, using the IAU approved conversion,

\begin{equation}
JY = (BY \times 0.999978641) + 0.041439661.
\end{equation}

\noindent The difference is slight, and given their published 
precision only 41 dates in the table have been changed.

\subsection{Zero-weighted Measures}

Measures not appearing in Table 3 and not used in the orbit solution
include those which are incomplete and list only the position angle 
and no separation (Herschel 1785, Struve 1837, Plummer 1878, Howe 
1879, Doberck 1896, 1902, Comstock 1906, Lohse 1908) as well as 
those which are measures of magnitude difference only (Pettit 1958, 
Kuiper 1950, Wieth-Knudsen 1957, Rakos et al.\ 1982). 

Others not included is the measure of Schembor (1939) which has an 
extremely large residual and appears to be a measure of the position
angle of the AB pair of this multiple system coupled to the 
separation of BC. Also not included is the measure of Van Biesbroeck
(1974). The residual is much larger than is typical for measures 
from this very experienced observer. In that paper, the measure of 
40 Eri BC in Table 1 is listed as having very small residuals to the
orbit of Wielen (1962)\footnote{This is, presumeably, Orbit 
III, as this was the preferred orbit in the 3$^{\it rd}$ Catalogue 
(Finsen \& Worley 1970), although four sets of elements are in 
Wielen (1962).}. 
However, there is either a typo in {\bf both} of the measures or 
there was a typo in the orbit residual. Given the ambiguity this 
mean position is not included. Had Van Biesbroeck been able to see 
the final manuscript to completion it would, no doubt, have been 
corrected. The measure of Chaname \& Gould (2004) has a very large 
difference in position angle from contemporaneous measures and is 
given an observation date of $``$approximately 2000" which is 
insufficiently precise for orbit determination and is also not 
included.

\section{The Orbit of 40 Eri BC}

Using the elements of Heintz (1974) to provide a first guess at the 
period, time of periastron passage and eccentricity, a method of 
differential correction was applied with the ``grid search" routine 
described in Hartkopf et al.\ (1989). Weights to the measures were 
applied using the methodology of Hartkopf et al.\ (2001). 
Briefly describing the weighting methodology the following 
factors were considered: telescope aperture, separation, number of
nights, and method of data acquisition. Arriving at the factors used
in weighting was accomplished by evaluating approximately 66,000 
observations of 450 well-characterized orbits in the generation of 
the orbit catalog Hartkopf et al.\ (2001). After performing the 
adaptive ``grid-search" until the step-size is very small rms values
are determined and weights adjusted. Measures made by micrometry are
zero-weighted when the residual is three times the rms. Measures 
made by photography or CCD have their weights reduced to 25\% of 
their previous value. The ``grid-search" is then repeated until 
lower tolerances in step size are met. These final weights are 
provided in Tables 2 \& 3.

Table 4 
lists the seven Campbell elements: $P$ (period, in years), $a$ 
(semi-major axis, in arcseconds), $i$ (inclination, in degrees), 
$\Omega$ (longitude of node, equinox 2000, in degrees), $T_0$ (epoch
of periastron passage, in fractional Julian year), $e$ 
(eccentricity), and $\omega$ (longtitude of periastron, in degrees).
Formal errors are listed with each element. Also provided in 
Table 4 are the parallax and mass ratio from van Leeuwen (2007) and 
Heintz (1974), respectively, used to determine their individual masses. 
This pair was identified by one of the authors (KNM) in Summer 2016 
as a pair suitable for orbit improvement and a preliminary version 
of these elements (determined without the measures of Table 2) 
appeared in the Commission G1 (n\'{e}e 26) Information Circular 
(Miles \& Mason 2017). For historical context, the earlier 
orbital elements of Heintz (1974; equinox 2000), Orbit III of Wielen
(1962; equinox unspecified), van den Bos (1926; equinox 1900) and 
Gore (1886; equinox 1880) are also given.

Figure 1 illustrates the new orbital solution, plotted together 
with all published data in the WDS database as well as the 
heretofore unpublished data in Table 2. In this figure, micrometric 
observations are indicated by plus signs, photographic measures by 
asterisks, adaptive optics by filled circles, CCD measures by
triangles and the four new measures from Table 2 as stars. ``$O-C$" 
lines connect each measure to its predicted position along the new 
orbit (shown as a thick solid line). Dashed ``$O-C$" lines indicate 
measures given zero weight in the final solution. A dot-dash line 
indicates the line of nodes, and a curved arrow in the lower right 
corner indicates the direction of orbital motion. The scale, in
arcseconds, is given on the left and bottom axis. Finally, the 
orbit of Heintz (1974) is shown as a dashed ellipse. 

Table 5 gives the ephemerides for the orbit over the years 2018 
through 2027, in annual increments. 

While the orbit has only completed 71\% of a full cycle, the orbit 
is quite well characterized. The criteria of Aitken (1935b):

\begin{quote}

... it is not worth while to compute the orbit of a double star
until the observed arc not only exceeds 180 degrees, but also defines
both ends of the apparent ellipse ...

\end{quote}

\noindent have been met. The orbit of Heintz (1974) lists no errors 
on the orbital elements which is reflected in his very low mass 
errors. That orbit was premature and appeared 22 years prior to 
reaching the northern limit of the orbit; this appears to be the 
primary reason for the incongruous mass solutions for these two 
stars. In addition to both ends of the apparent ellipse now being 
well-characterized, a more accurate and precise parallax 
(200.62$\pm$0.23 mas, van Leeuwen 2007) has been determined and the 
number of measures has increased by 14\%. Note that the 
parallax is for the primary of the physical multiple. If we assume 
the AB mean motion of 0.026 $^{\circ}$/yr is representative, then 
the parallax difference for BC would be quite close to this value 
and within 0.066\%. While SIMBAD lists 198.24 mas for B (Holberg et 
al.\ 2002) this corresponds to the original Hipparcos solution (ESA 
1997) for A. We use this re-reduction of the Hipparcos value. The 
orbit has very small errors of 0.7\% in the semimajor axis (a$''$) 
and 0.3\% in the period (P), yielding an error of 3.1\% in the mass 
sum. The mass sum, \massAB ~is 0.776$\pm$0.024 \msun . Using the 
mass ratio from Heintz (1974) gives individual masses of 
0.575$\pm$0.018 \msun ~for the white dwarf and 0.2041$\pm$0.0064 
\msun ~for the M dwarf companion. 

The newly determined mass for the M dwarf companion falls within the
1$\sigma$ error of its value in Henry et al.\ (1999) of 
0.177$\pm$0.029 \msun. The mass error here is comparable to the 
other mass errors of Table 11 of Benedict et al.\ (2016). The mass 
is less than that of GJ 791.2, also classified as M4.5V, determined 
in Benedict et al.\ (see Tables 2 \& 10).

If the solution presented in Table 4 is representative of the true 
motion and we were to wait two more observing seasons and observe 
the pair monthly, when accessible, the resulting errors would 
improve less than a tenth of a percent in P, a$''$ or \massAB . The 
most significant improvements could occur with data obtained as it 
approaches the next periastron passage (predicted for 2077.7) or 
when the system has been observed for a complete revolution 
(predicted for 2081.2). Due to the geometry of the system, the 
closest approach of 1\farcs32 is predicted to occur more than a 
decade before periastron : 2066.2. 

With the post-AGB mass loss of the B component of the system, 
the orbital elements must have gone through significant evolution. 
Zhao et al.\ (2011) determine ages of A and C through analysis of
chromospheric activity of $5.0^{6.1}_{4.0}$ Gyr and an age of 
 $4.9^{6.0}_{3.9}$ Gyr for B based on the evolutionary lifetime
of the progenitor plus cooling time. Sousa et al.\ (2008) determine
metallicity of A as $-0.31\pm0.03$. These two accurate and precise
results coupled with the very accurate and precise masses determined
here, will help enable study of the complicated interplay between 
mass, age and metallicity of all three components in this 
hierarchical multiple.

%
%
%
%
%
%
%
%
%

\begin{figure}[p]
\begin{center}
{\epsfxsize 5.5in \epsffile{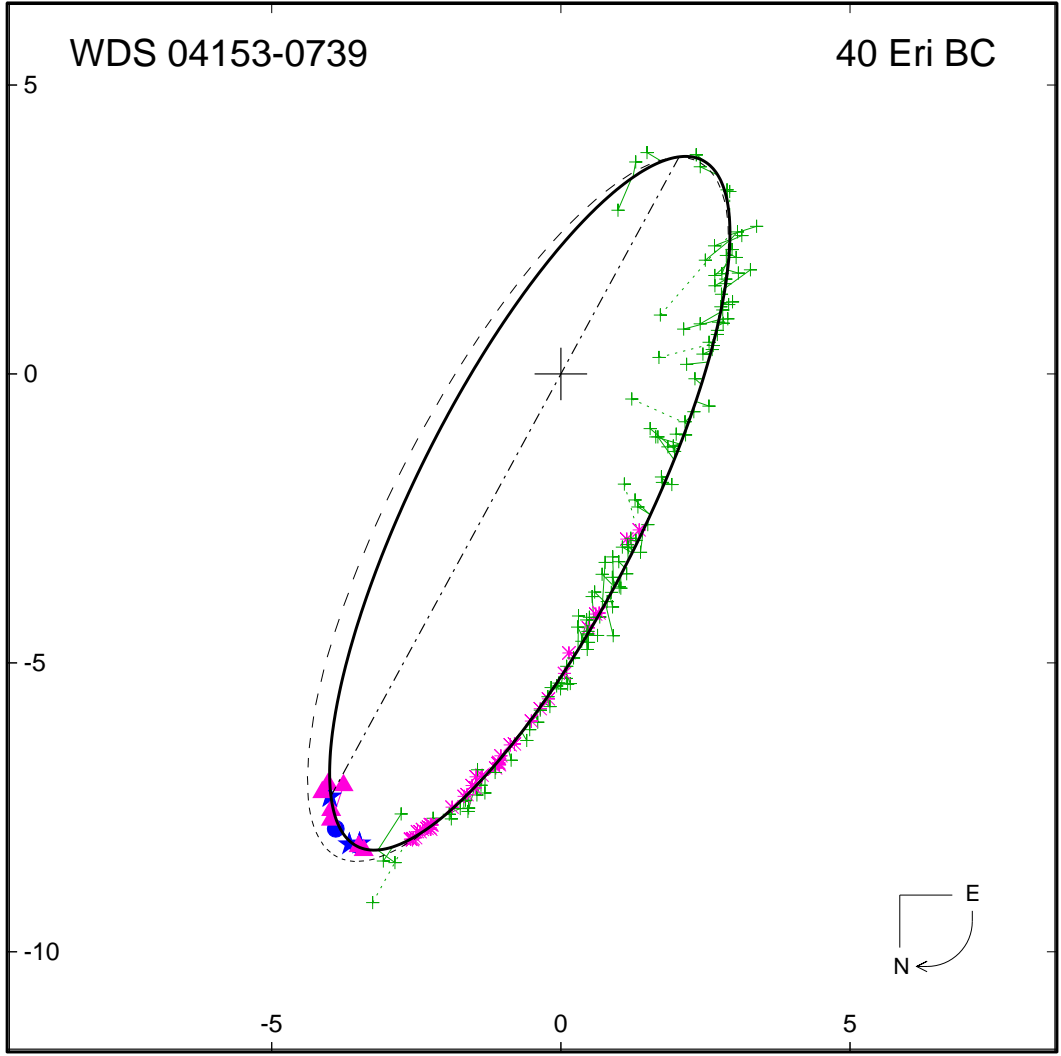}}
\vskip 0.05in
\end{center}
\caption{\small New orbit of 40 Eri BC as described in the text. The
solid curve is the solution presented in Table 4. The dashed curve 
is the orbit of Heintz (1974). The zero-weighted and aberrant 
measures of Schembor (1939), Van Biesbroeck (1974), and Chaname \& 
Gould (2004) are not plotted for cosmetic reasons.}
\end{figure}

In addition to determining a mass for the red dwarf, the value of
0.575$\pm$0.018 \msun ~for the white dwarf is now in agreement with
those determined using the gravitational redshift (for example,  
within 1$\sigma$ of the result 0.53$\pm$0.04 \msun ~from Koester \& 
Weidemann 1991). While the results match well here, it is 
unclear if they agree well-enough to make one determination 
redundant. For example, in the case of Sirius B the results are 
slightly discrepant with a dynamical mass of 1.018$\pm$0.011 \msun 
(Bond et al.\ 2017) and a mass from the gravitational redshift of 
0.978$\pm$0.005 \msun (Barstow et al.\ 2005).

Now that the mass from the orbit matches that from the gravitational
redshift, this source of consternation has gone away and it is not 
necessary to invoke other more exotic solutions to the problem. 
Patience is a virtue.

\acknowledgments

Howard Bond who suggested examining the object and publishing the
orbit now rather than waiting on more data is gratefully thanked.
The best is the enemy of the good. The referee is heartily thanked 
for many helpful suggestions. This publication makes use of data 
products from the Two Micron All Sky Survey, which is a joint 
project of the University of Massachusetts and the Infrared 
Processing and Analysis Center/California Institute of Technology, 
funded by the National Aeronautics and Space Administration and 
the National Science Foundation. This research has also made use 
of the SIMBAD database, operated at CDS, Strasbourg, France and 
NASA's Astrophysics Data System. Thanks are extended to Brian 
Luzum and the U.S.\ Naval Observatory for their continued support 
of the Double Star Program.

\begin{deluxetable}{ccccc}

\tablenum{1}
\tablewidth{0pt}
\tablecaption{40 Eri Component Properties}
\tablehead{
\colhead{~} &
\colhead{A} & 
\colhead{B} & 
\colhead{C} &
\colhead{Source} \\
}
\startdata
$\alpha$ (2000) & \phs04\phn15\phn16.32    &                          &                & van Leeuwen 2007             \\
                &                          & \phs04\phn15\phn21.79    &                & Zacharias et al.\ 2003       \\
                &                          &                          & Table 5        &                              \\
$\delta$ (2000) &  $-$07\phn39\phn10.3\phn &                          &                & van Leeuwen 2007             \\
                &                          &  $-$07\phn39\phn29.1\phn &                & Zacharias et al.\ 2003       \\
                &                          &                          & Table 5        &                              \\
$\mu_{\alpha}$  & $-$2240.12 mas/yr        &                          &                & van Leeuwen 2007             \\
                &                          & $-$2228.3 mas/yr         &                & Zacharias et al.\ 2003       \\
                &                          &                          & $-$2239 mas/yr & Salim \& Gould 2003          \\
$\mu_{\delta}$  & $-$3420.27 mas/yr        &                          &                & van Leeuwen 2007             \\
                &                          & $-$3377.1 mas/yr         &                & Zacharias et al.\ 2003       \\
                &                          &                          & $-$3419 mas/yr & Salim \& Gould 2003          \\
Parallax        & 200.62 mas               & ~                        & ~              & van Leeuwen 2007             \\
Spectral type   & K0.5V                    &                          &                & Gray et al.\ 2006            \\
                &                          & DA2.9                    &                & Gianninas et al.\ 2011       \\
                &                          &                          & M4.5V          & Alonso-Floriano et al.\ 2015 \\
V mag           & 4.43                     &                          &                & Ducati 2002                  \\
                &                          & 9.53                     &                & Kidder et al.\ 1991          \\
                &                          &                          & 11.17          & Holberg et al.\ 2012         \\
\enddata
\end{deluxetable}

\begin{deluxetable}{cccccccccl}

\tablenum{2}
\tablewidth{0pt}
\tablecaption{New Measurements of 40 Eri BC}
\tablehead{
\colhead{Julian} &
\colhead{$\theta$} & 
\colhead{$\sigma\theta$} & 
\colhead{$\rho$} &
\colhead{$\sigma\rho$} &
\colhead{n} &
\colhead{O$-$C} &
\colhead{O$-$C} &
\colhead{Weight} &
\colhead{Source} \\
\colhead{Epoch} & 
\colhead{($\circ$)} & 
\colhead{($\circ$)} & 
\colhead{($''$)} &
\colhead{($''$)} &
\colhead{~} &
\colhead{($\circ$)} & 
\colhead{($''$)} &
\colhead{~} &
\colhead{~} \\
}
\startdata
1998.87\phn\phn & 336.8 &     & 8.84\phn &       & 1 & \phs0.6 &  $-$0.052 & 20.0 & 2MASS$^1$    \\
1999.97\phn\phn & 335.8 & 0.1 & 8.924    & 0.011 & 4 &  $-$0.1 & \phs0.042 & 40.0 & UCAC4$^2$    \\
2017.1322       & 331.5 & 0.0 & 8.334    & 0.017 & 2 & \phs0.2 & \phs0.060 & 28.3 & USNO Speckle \\
2017.1901       & 331.4 & 0.0 & 8.337    & 0.007 & 4 & \phs0.1 & \phs0.067 & 40.0 & USNO Speckle \\
\enddata
\tablenotetext{~}{1 : Cutri et al.\ (2003), All-sky Release. See {\tt Vizier On-line Data Catalog: II/246}.}
\tablenotetext{~}{2 : Zacharias et al.\ 2013}

\end{deluxetable}

\begin{deluxetable}{ccccccclc}

\tablenum{3}
\tablewidth{0pt}
\tablecaption{Catalog Measurements of 40 Eri BC}
\tablehead{
\colhead{Julian} &
\colhead{$\theta$} & 
\colhead{$\rho$} &
\colhead{n} &
\colhead{O$-$C} &
\colhead{O$-$C} &
\colhead{Weight} &
\colhead{Source} &
\colhead{Notes} \\
\colhead{Epoch} & 
\colhead{($\circ$)} & 
\colhead{($''$)} &
\colhead{~} &
\colhead{($\circ$)} & 
\colhead{($''$)} &
\colhead{~} &
\colhead{~} &
\colhead{~} \\
}
\startdata
1851.06\phn\phn &       160.0\phn    & 3.\phn\phn\phn & \phn1 & \phs0.6 &  $-$0.599 & \phn1.1 & Dawes 1867                 &      \\
1851.22\phn\phn &       159.8\phn    & 3.89\phn       & \phn4 & \phs0.7 & \phs0.269 & \phn4.3 & Struve 1878                &      \\
1855.06\phn\phn &       158.0\phn    & 4.11\phn       & \phn6 & \phs4.1 & \phs0.040 & \phn5.6 & Struve 1878                &      \\
1864.84\phn\phn &       147.6\phn    & 4.46\phn       & \phn2 & \phs4.3 & \phs0.032 & \phn3.5 & Struve 1878                &      \\
1864.85\phn\phn &       147.61       & 4.455          & \phn2 & \phs4.3 & \phs0.027 & \phn2.1 & Winnecke 1869              &      \\
1866.96\phn\phn &       145.4\phn    & 4.32\phn       & \phn3 & \phs4.3 &  $-$0.062 & \phn4.1 & Struve 1878                &      \\
1873.85\phn\phn &       137.3\phn    & 4.29\phn       & \phn5 & \phs4.1 & \phs0.262 & \phn5.3 & Struve 1878                &      \\
1875.90\phn\phn &       136.6\phn    & 4.3\phn\phn    & \phn1 & \phs6.1 & \phs0.420 & \phn0.0 & Lewis 1906                 & A,B  \\
1877.12\phn\phn &       126.4\phn    & 4.24\phn       & \phn2 &  $-$2.4 & \phs0.453 & \phn3.3 & Struve 1893                &      \\
1877.12\phn\phn &       120.0\phn    & 2.\phn\phn\phn & \phn1 &  $-$8.8 &  $-$1.787 & \phn0.0 & Flammarion 1878            & A,C  \\
1877.79\phn\phn &       129.2\phn    & 3.46\phn       & \phn2 & \phs1.3 &  $-$0.274 & \phn1.9 & Stone 1878                 & D    \\
1877.86\phn\phn &       128.2\phn    & 3.92\phn       & \phn7 & \phs0.4 & \phs0.192 &    16.3 & Burnham 1879               &      \\
1877.87\phn\phn &       127.6\phn    & 3.18\phn       & \phn2 &  $-$0.2 &  $-$0.547 & \phn2.9 & Howe 1878                  &      \\
1877.95\phn\phn &       126.8\phn    & 3.94\phn       & \phn4 &  $-$0.8 & \phs0.219 & \phn7.0 & Dembowski 1884             &      \\
1879.05\phn\phn &       125.4\phn    & 3.66\phn       & \phn4 &  $-$0.6 & \phs0.029 &    11.4 & Burnham 1883               &      \\
1879.181\phn    &       125.0\phn    & 3.52\phn       & \phn2 &  $-$0.8 &  $-$0.101 &    11.6 & Hall 1877                  & E    \\
1879.68\phn\phn &       123.0\phn    & 3.64\phn       & \phn2 &  $-$2.0 & \phs0.060 & \phn3.7 & Burnham 1887               &      \\
1879.75\phn\phn &       119.3\phn    & 3.29\phn       & \phn1 &  $-$5.6 &  $-$0.284 & \phn1.3 & Stone 1882                 &      \\
1880.09\phn\phn &       121.3\phn    & 3.28\phn       & \phn5 &  $-$3.0 &  $-$0.266 &    11.5 & Burnham 1883               &      \\
1880.95\phn\phn &       122.0\phn    & 3.16\phn       & \phn5 &  $-$0.9 &  $-$0.314 &    11.1 & Burnham 1883               &      \\
1881.84\phn\phn &       119.0\phn    & 3.53\phn       & \phn6 &  $-$2.4 & \phs0.131 &    13.5 & Burnham 1883               &      \\
1882.119\phn    &       118.2\phn    & 3.24\phn       & \phn2 &  $-$2.7 &  $-$0.135 &    10.3 & Hall 1892                  & E    \\
1883.00\phn\phn &       119.2\phn    & 3.07\phn       & \phn2 &  $-$0.1 &  $-$0.231 & \phn6.8 & Burnham 1883               &      \\
1883.807\phn    &       115.8\phn    & 3.10\phn       & \phn2 &  $-$1.9 &  $-$0.133 & \phn9.8 & Hall 1892                  & E    \\
1884.16\phn\phn &       118.2\phn    & 3.74\phn       & \phn1 & \phs1.2 & \phs0.536 & \phn2.1 & Struve 1893                &      \\
1886.00\phn\phn &       111.9\phn    & 3.14\phn       & \phn2 &  $-$1.3 & \phs0.087 & \phn9.6 & Leavenworth \& Beal 1930   &      \\
1886.002\phn    &       112.2\phn    & 3.22\phn       & \phn3 &  $-$1.0 & \phs0.167 &    12.1 & Leavenworth \& Muller 1915 & E    \\
1886.095\phn    &       112.2\phn    & 3.00\phn       & \phn6 &  $-$0.8 &  $-$0.045 &    16.2 & Hall 1892                  & E    \\
1886.92\phn\phn &       111.0\phn    & 3.01\phn       & \phn3 &  $-$0.1 & \phs0.031 & \phn2.6 & Tarrant 1889               &      \\
1887.14\phn\phn &       109.2\phn    & 2.56\phn       & \phn4 &  $-$1.4 &  $-$0.402 & \phn8.7 & Schiaparelli 1909          &      \\
1888.08\phn\phn &       109.5\phn    & 2.26\phn       & \phn2 & \phs1.1 &  $-$0.630 & \phn5.7 & Schiaparelli 1909          &      \\
1888.121\phn    &       107.7\phn    & 3.04\phn       & \phn5 &  $-$0.6 & \phs0.153 &    15.1 & Hall 1892                  & E    \\
1888.84\phn\phn &       106.8\phn    & 2.94\phn       & \phn3 & \phs0.3 & \phs0.107 & \phn5.7 & Burnham 1894               &      \\
1888.870\phn    &       105.0\phn    & 2.81\phn       & \phn3 &  $-$1.4 &  $-$0.021 & \phn2.4 & Tarrant 1890               & E    \\
1889.03\phn\phn &       107.6\phn    & 2.87\phn       & \phn2 & \phs1.6 & \phs0.050 & \phn6.7 & Schiaparelli 1909          &      \\
1889.123\phn    &       103.6\phn    & 2.79\phn       & \phn4 &  $-$2.2 &  $-$0.023 &    12.2 & Hall 1892                  & E    \\
1890.73\phn\phn &       100.0\phn    & 2.68\phn       & \phn4 &  $-$1.5 &  $-$0.022 &    15.9 & Burnham 1894               &      \\
1890.98\phn\phn &    \phn99.0\phn    & 1.72\phn       & \phn3 &  $-$1.8 &  $-$0.966 & \phn0.0 & Hough 1894                 & A    \\
1891.00\phn\phn &       101.5\phn    & 2.62\phn       & \phn2 & \phs0.8 &  $-$0.064 & \phn6.3 & Schiaparelli 1909          &      \\
1891.056\phn    &    \phn98.6\phn    & 2.65\phn       & \phn5 &  $-$1.9 &  $-$0.031 &    12.9 & Hall 1892                  & E    \\
1891.78\phn\phn &    \phn97.4\phn    & 2.48\phn       & \phn4 &  $-$1.0 &  $-$0.156 &    14.4 & Burnham 1894               &      \\
1893.212\phn    &    \phn93.8\phn    & 2.18\phn       & \phn1 &  $-$0.3 &  $-$0.375 & \phn2.6 & Comstock 1896              & E    \\
1895.912\phn    &    \phn87.4\phn    & 2.32\phn       & \phn2 & \phs2.2 &  $-$0.116 & \phn1.4 & Collins 1896               & E    \\
1897.97\phn\phn &    \phn77.2\phn    & 2.62\phn       & \phn3 &  $-$0.8 & \phs0.239 & \phn4.0 & Aitken 1914                &      \\
1899.11\phn\phn &    \phn73.6\phn    & 2.39\phn       & \phn2 &  $-$0.2 & \phs0.025 & \phn2.9 & Aitken 1914                &      \\
1899.803\phn    &    \phn68.4\phn    & 2.30\phn       & \phn3 &  $-$2.9 &  $-$0.061 & \phn4.4 & Doolittle 1905             & E    \\
1900.743\phn    &    \phn70.\phn\phn & 1.3\phn\phn    & \phn1 & \phs2.2 &  $-$1.061 & \phn0.0 & Comas Sola 1900            & A,C,E  \\
1900.926\phn    &    \phn63.4\phn    & 2.40\phn       & \phn2 &  $-$3.8 & \phs0.038 & \phn3.7 & Doolittle 1905             & E    \\
1902.002\phn    &    \phn61.9\phn    & 2.25\phn       & \phn4 &  $-$1.4 &  $-$0.123 & \phn5.3 & Comstock 1906              & E    \\
1903.142\phn    &    \phn55.2\phn    & 2.24\phn       & \phn4 &  $-$4.0 &  $-$0.156 & \phn5.0 & Doolittle 1905             & E    \\
1903.183\phn    &    \phn55.9\phn    & 1.97\phn       & \phn1 &  $-$3.1 &  $-$0.427 & \phn2.3 & Comstock 1906              & E    \\
1903.87\phn\phn &    \phn56.8\phn    & 2.31\phn       & \phn2 & \phs0.2 &  $-$0.106 & \phn7.4 & Aitken 1914                &      \\
1904.105\phn    &    \phn58.0\phn    & 1.81\phn       & \phn1 & \phs2.2 &  $-$0.613 & \phn2.2 & Doolittle 1905             & E    \\
1904.70\phn\phn &    \phn55.2\phn    & 2.38\phn       & \phn3 & \phs1.5 &  $-$0.063 &    13.6 & Burnham 1906               &      \\
1905.11\phn\phn &    \phn56.5\phn    & 2.00\phn       & \phn1 & \phs4.2 &  $-$0.459 & \phn0.2 & Lohse 1908                 & C    \\
1907.80\phn\phn &    \phn43.8\phn    & 2.49\phn       & \phn4 & \phs0.1 &  $-$0.101 &    16.7 & Burnham 1913               &      \\
1907.97\phn\phn &    \phn44.6\phn    & 2.71\phn       & \phn2 & \phs1.4 & \phs0.109 & \phn4.4 & Wirtz 1912                 &      \\
1908.83\phn\phn &    \phn42.6\phn    & 2.57\phn       & \phn5 & \phs2.0 &  $-$0.083 &    19.5 & Burnham 1913               &      \\
1912.04\phn\phn &    \phn29.6\phn    & 2.66\phn       & \phn2 &  $-$2.4 &  $-$0.225 & \phn8.9 & Aitken 1914                &      \\
1912.11\phn\phn &    \phn30.0\phn    & 2.53\phn       & \phn1 &  $-$1.9 &  $-$0.361 & \phn1.3 & Vanderdonck 1912           & F    \\
1913.138\phn    &    \phn29.5\phn    & 3.01\phn       & \phn3 & \phs0.1 & \phs0.035 & \phn5.0 & Van Biesbroeck 1920        & E    \\
1914.063\phn    &    \phn23.6\phn    & 3.38\phn       & \phn1 &  $-$3.8 & \phs0.326 & \phn3.2 & Van Biesbroeck 1920        & E    \\
1915.09\phn\phn &    \phn29.5\phn    & 2.2\phn\phn    & \phn1 & \phs4.3 &  $-$0.946 & \phn0.0 & Rabe 1923                  & A,C  \\
1915.13\phn\phn &    \phn26.2\phn    & 3.02\phn       & \phn5 & \phs1.1 &  $-$0.129 &    12.5 & Olivier 1920               &      \\
1915.64\phn\phn &    \phn21.30       & 3.077          & \phn8 &  $-$2.8 &  $-$0.119 &    16.6 & Heintz 1974                &      \\
1915.860\phn    &    \phn23.8\phn    & 3.16\phn       & \phn3 & \phs0.2 &  $-$0.056 &    12.0 & Van Biesbroeck 1927        & E    \\
1916.83\phn\phn &    \phn20.9\phn    & 3.22\phn       & \phn3 &  $-$0.9 &  $-$0.087 & \phn4.8 & Olivier 1917               &      \\
1917.08\phn\phn &    \phn19.1\phn    & 3.18\phn       & \phn1 &  $-$2.2 &  $-$0.151 & \phn7.6 & Aitken 1923                &      \\
1917.16\phn\phn &    \phn22.6\phn    & 3.09\phn       & \phn2 & \phs1.4 &  $-$0.248 & \phn4.6 & Comstock 1921              &      \\
1918.14\phn\phn &    \phn20.9\phn    & 3.17\phn       & \phn3 & \phs1.5 &  $-$0.263 & \phn5.7 & Comstock 1921              &      \\
1919.09\phn\phn &    \phn16.7\phn    & 3.40\phn       & \phn3 &  $-$1.1 &  $-$0.127 & \phn6.0 & Leavenworth \& Beal 1930   &      \\
1920.008\phn    &    \phn17.8\phn    & 3.64\phn       & \phn3 & \phs1.5 & \phs0.022 & \phn7.3 & Bernewitz 1962             &      \\
1920.132\phn    &    \phn15.2\phn    & 3.85\phn       & \phn3 &  $-$1.0 & \phs0.219 &    13.2 & Pavel 1962                 &      \\
1921.134\phn    &    \phn14.9\phn    & 3.82\phn       & \phn5 & \phs0.3 & \phs0.088 &    10.1 & Bernewitz 1962             &      \\
1921.516\phn    &    \phn13.8\phn    & 3.63\phn       & \phn3 &  $-$0.3 &  $-$0.141 &    12.0 & Van Biesbroeck 1927        &      \\
1921.79\phn\phn &    \phn11.2\phn    & 3.54\phn       & \phn2 &  $-$2.5 &  $-$0.259 &    10.8 & Aitken 1923                &      \\
1922.00\phn\phn &    \phn15.4\phn    & 3.29\phn       & \phn2 & \phs2.0 &  $-$0.531 & \phn2.1 & Abetti 1922                &      \\
1922.02\phn\phn &    \phn12.6\phn    & 3.88\phn       & \phn2 &  $-$0.8 & \phs0.057 & \phn1.9 & Nechvile 1924              &      \\
1922.988\phn    &    \phn11.1\phn    & 4.02\phn       & \phn2 &  $-$1.0 & \phs0.097 & \phn5.5 & Dick 1962                  &      \\
1923.010\phn    &    \phn12.1\phn    & 4.13\phn       & \phn2 & \phs0.1 & \phs0.205 &    12.3 & Struve 1926                &      \\
1924.07\phn\phn & \phn\phn8.4\phn    & 3.82\phn       & \phn2 &  $-$2.3 &  $-$0.216 &    10.8 & Aitken 1927                &      \\
1924.142\phn    &    \phn10.9\phn    & 4.62\phn       & \phn2 & \phs0.3 & \phs0.576 & \phn5.5 & Dick 1962                  &      \\
1925.02\phn\phn &    \phn12.8\phn    & 3.35\phn       & \phn5 & \phs3.3 &  $-$0.786 &    10.1 & van den Bos 1925           &      \\
1925.87\phn\phn & \phn\phn7.87       & 4.200          & \phn3 &  $-$0.7 &  $-$0.025 &    22.5 & Heintz 1974                &      \\
1926.19\phn\phn & \phn\phn8.7\phn    & 4.26\phn       & \phn5 & \phs0.5 & \phs0.001 &    26.9 & van den Bos 1928           &      \\
1926.65\phn\phn & \phn\phn8.7\phn    & 4.19\phn       & \phn1 & \phs1.0 &  $-$0.118 &    13.0 & Alden 1936                 &      \\
1927.06\phn\phn & \phn\phn7.6\phn    & 3.89\phn       & \phn6 & \phs0.4 &  $-$0.461 & \phn7.9 & Rabe 1930                  &      \\
1928.07\phn\phn & \phn\phn5.8\phn    & 4.42\phn       & \phn1 &  $-$0.4 &  $-$0.038 &    13.3 & Alden 1936                 &      \\
1928.15\phn\phn & \phn\phn6.3\phn    & 4.24\phn       & \phn1 & \phs0.2 &  $-$0.226 & \phn4.0 & van den Bos 1928           &      \\
1928.16\phn\phn & \phn\phn5.6\phn    & 4.48\phn       & \phn1 &  $-$0.5 & \phs0.013 & \phn4.2 & van den Bos 1928           &      \\
1928.90\phn\phn & \phn\phn7.6\phn    & 4.57\phn       & \phn3 & \phs2.3 & \phs0.024 &    13.3 & Voute 1932                 &      \\
1929.04\phn\phn & \phn\phn5.1\phn    & 4.79\phn       & \phn1 &  $-$0.1 & \phs0.230 &    13.3 & Finsen 1929                &      \\
1929.04\phn\phn & \phn\phn5.4\phn    & 4.52\phn       & \phn2 & \phs0.2 &  $-$0.040 &    17.0 & van den Bos 1929           &      \\
1929.04\phn\phn & \phn\phn5.4\phn    & 4.54\phn       & \phn1 & \phs0.2 &  $-$0.020 &    13.3 & Finsen 1929                &      \\
1929.56\phn\phn & \phn\phn3.4\phn    & 4.39\phn       & \phn2 &  $-$1.3 &  $-$0.226 &    10.8 & Aitken 1935a               &      \\
1929.72\phn\phn & \phn\phn5.5\phn    & 4.67\phn       & \phn3 & \phs1.0 & \phs0.037 &    13.3 & Voute 1932                 &      \\
1930.60\phn\phn & \phn\phn5.6\phn    & 4.28\phn       & \phn4 & \phs1.9 &  $-$0.446 & \phn3.7 & Wamer 1932                 &      \\
1930.82\phn\phn & \phn\phn4.1\phn    & 4.64\phn       & \phn3 & \phs0.6 &  $-$0.109 & \phn4.7 & Wallenquist 1934           &      \\
1931.18\phn\phn & \phn\phn3.8\phn    & 4.20\phn       & \phn3 & \phs0.6 &  $-$0.587 & \phn2.0 & Baize \& Igounet 1930      &      \\
1932.26\phn\phn & \phn\phn1.3\phn    & 4.83\phn       & \phn1 &  $-$1.0 &  $-$0.071 &    13.9 & Alden 1936                 &      \\
1932.70\phn\phn & \phn\phn2.2\phn    & 4.92\phn       & \phn4 & \phs0.3 &  $-$0.027 &    15.3 & Voute 1932                 &      \\
1934.16\phn\phn & \phn\phn0.9\phn    & 5.37\phn       & \phn6 & \phs0.2 & \phs0.270 & \phn9.3 & Baize 1935                 &      \\
1934.45\phn\phn & \phn\phn1.4\phn    & 5.36\phn       & \phn7 & \phs0.9 & \phs0.230 &    10.1 & Baize 1942                 &      \\
1934.97\phn\phn &       359.6\phn    & 5.45\phn       & \phn4 &  $-$0.5 & \phs0.266 & \phn2.8 & Inaba 1935                 &      \\
1935.04\phn\phn & \phn\phn0.7\phn    & 5.27\phn       & \phn4 & \phs0.7 & \phs0.079 &    24.1 & van den Bos 1935           &      \\
1935.31\phn\phn & \phn\phn0.34       & 5.182          &    10 & \phs0.5 &  $-$0.037 &    44.3 & Heintz 1974                &      \\
1936.03\phn\phn & \phn\phn0.8\phn    & 5.06\phn       & \phn2 & \phs1.5 &  $-$0.234 & \phn7.6 & Rabe 1939                  &      \\
1936.85\phn\phn &       358.9\phn    & 5.40\phn       & \phn4 & \phs0.2 & \phs0.022 &    11.8 & Simonov 1951               &      \\
1938.15\phn\phn &       359.3\phn    & 5.35\phn       & \phn2 & \phs1.5 &  $-$0.161 & \phn8.0 & Rabe 1939                  &      \\
1938.76\phn\phn &       357.9\phn    & 5.43\phn       & \phn4 & \phs0.5 &  $-$0.142 &    15.3 & Voute 1947                 &      \\
1939.35\phn\phn &       357.49       & 5.627          & \phn8 & \phs0.4 &  $-$0.005 &    39.2 & Heintz 1974                &      \\
1939.93\phn\phn &       357.8\phn    & 5.76\phn       & \phn5 & \phs1.1 & \phs0.070 & \phn9.1 & Baize 1942                 &      \\
1941.26\phn\phn &       356.17       & 5.804          & \phn8 & \phs0.3 &  $-$0.018 &    39.6 & Heintz 1974                &      \\
1942.05\phn\phn &       355.9\phn    & 6.04\phn       & \phn4 & \phs0.5 & \phs0.140 &    24.1 & van den Bos 1948           &      \\
1942.12\phn\phn &       357.4\phn    & 5.59\phn       & \phn2 & \phs2.0 &  $-$0.317 & \phn8.4 & Rabe 1953                  &      \\
1942.76\phn\phn &       356.2\phn    & 5.83\phn       & \phn3 & \phs1.2 &  $-$0.140 &    10.5 & Voute 1955                 &      \\
1943.14\phn\phn &       354.7\phn    & 6.18\phn       & \phn3 &  $-$0.1 & \phs0.174 &    11.3 & Rabe 1953                  &      \\
1943.88\phn\phn &       354.85       & 6.028          &    14 & \phs0.5 &  $-$0.050 &    52.4 & Heintz 1974                &      \\
1945.53\phn\phn &       354.4\phn    & 6.37\phn       & \phn5 & \phs0.9 & \phs0.135 &    10.1 & Baize 1948                 &      \\
1948.12\phn\phn &       352.58       & 6.454          &    10 & \phs0.4 &  $-$0.022 &    44.3 & Heintz 1974                &      \\
1948.40\phn\phn &       351.97       & 6.479          & \phn4 &  $-$0.1 &  $-$0.022 &    28.0 & Heintz 1974                &      \\
1949.00\phn\phn &       352.4\phn    & 6.74\phn       & \phn2 & \phs0.6 & \phs0.185 &    17.0 & van den Bos 1951           &      \\
1950.71\phn\phn &       350.83       & 6.690          & \phn6 &  $-$0.2 &  $-$0.018 &    34.3 & Heintz 1974                &      \\
1951.733\phn    &       350.65       & 6.809          & \phn1 & \phs0.1 & \phs0.012 &    14.0 & The 1970                   &      \\
1951.812\phn    &       350.70       & 6.823          & \phn1 & \phs0.2 & \phs0.019 &    14.0 & The 1970                   &      \\
1951.829\phn    &       350.61       & 6.832          & \phn1 & \phs0.1 & \phs0.027 &    14.0 & The 1970                   &      \\
1951.886\phn    &       350.84       & 6.847          & \phn1 & \phs0.4 & \phs0.037 &    14.0 & The 1970                   &      \\
1952.89\phn\phn &       350.37       & 6.911          &    10 & \phs0.3 & \phs0.015 &    44.3 & Heintz 1974                &      \\
1953.99\phn\phn &       350.4\phn    & 6.99\phn       & \phn3 & \phs0.8 & \phs0.002 &    20.8 & van den Bos 1956           &      \\
1955.13\phn\phn &       349.5\phn    & 7.37\phn       & \phn4 & \phs0.4 & \phs0.287 &    18.7 & Worley 1956                &      \\
1955.18\phn\phn &       348.67       & 7.077          & \phn4 &  $-$0.4 &  $-$0.010 &    28.0 & Heintz 1974                &      \\
1955.84\phn\phn &       348.5\phn    & 7.43\phn       & \phn4 &  $-$0.3 & \phs0.290 &    18.9 & Worley 1957                &      \\
1956.855\phn    &       347.90       & 7.124          & \phn1 &  $-$0.5 &  $-$0.097 &    14.0 & The 1970                   &      \\
1957.73\phn\phn &       348.8\phn    & 7.25\phn       & \phn3 & \phs0.7 &  $-$0.040 &    20.8 & van den Bos 1959           &      \\
1957.748\phn    &       348.10       & 7.315          & \phn1 & \phs0.0 & \phs0.024 &    14.0 & The 1970                   &      \\
1957.88\phn\phn &       347.8\phn    & 7.74\phn       & \phn4 &  $-$0.2 & \phs0.439 &    20.0 & Worley 1960                &      \\
1957.890\phn    &       347.62       & 7.286          & \phn1 &  $-$0.4 &  $-$0.016 &    14.0 & The 1970                   &      \\
1958.07\phn\phn &       347.9\phn    & 7.00\phn       & \phn3 &  $-$0.1 &  $-$0.316 &    24.8 & Couteau 1958               &      \\
1959.51\phn\phn &       347.8\phn    & 7.68\phn       & \phn2 & \phs0.4 & \phs0.254 &    17.0 & van den Bos 1960           &      \\
1959.852\phn    &       347.23       & 7.460          & \phn1 &  $-$0.1 & \phs0.009 &    14.0 & Kamper 1976                &      \\
1959.92\phn\phn &       347.0\phn    & 7.58\phn       & \phn4 &  $-$0.3 & \phs0.123 &    25.3 & Worley 1962                &      \\
1960.62\phn\phn &       346.91       & 7.499          & \phn6 &  $-$0.1 &  $-$0.010 &    34.3 & Heintz 1974                &      \\
1961.86\phn\phn &       347.2\phn    & 7.56\phn       & \phn4 & \phs0.6 &  $-$0.038 &    24.1 & van den Bos 1962           &      \\
1963.92\phn\phn &       345.73       & 7.729          & \phn4 &  $-$0.2 &  $-$0.012 &    28.0 & Heintz 1974                &      \\
1964.710\phn    &       345.6\phn    & 7.86\phn       & \phn1 &  $-$0.0 & \phs0.066 &    12.6 & Worley 1971                & E    \\
1965.04\phn\phn &       346.0\phn    & 7.93\phn       & \phn3 & \phs0.5 & \phs0.114 &    20.8 & van den Bos 1966           &      \\
1965.96\phn\phn &       346.8\phn    & 7.72\phn       & \phn2 & \phs1.6 &  $-$0.156 & \phn7.9 & Newburg 1967               &      \\
1968.01\phn\phn &       343.8\phn    & 8.00\phn       & \phn1 &  $-$0.8 &  $-$0.004 & \phn9.4 & Knipe 1969                 &      \\
1969.883\phn    &       343.89       & 8.116          & \phn1 &  $-$0.1 & \phs0.002 &    14.0 & Kamper 1976                & E    \\
1969.97\phn\phn &       343.70       & 8.117          & \phn8 &  $-$0.3 &  $-$0.002 &    39.6 & Heintz 1974                &      \\
1970.006\phn    &       343.89       & 8.120          & \phn1 &  $-$0.1 &  $-$0.001 &    14.0 & Kamper 1976                & E    \\
1970.733\phn    &       343.79       & 8.174          & \phn1 & \phs0.1 & \phs0.012 &    14.0 & Josties et al.\ 1974       & E    \\
1970.763\phn    &       343.57       & 8.170          & \phn1 &  $-$0.1 & \phs0.006 &    14.0 & Josties et al.\ 1974       & E    \\
1970.93\phn\phn &       343.46       & 8.198          & \phn9 &  $-$0.2 & \phs0.025 &    42.0 & Heintz 1974                &      \\
1971.047\phn    &       343.92       & 8.195          & \phn1 & \phs0.3 & \phs0.015 &    14.0 & Josties et al.\ 1974       & E    \\
1972.00\phn\phn &       342.97       & 8.238          & \phn6 &  $-$0.4 & \phs0.007 &    34.3 & Heintz 1974                &      \\
1972.97\phn\phn &       342.78       & 8.288          & \phn6 &  $-$0.3 & \phs0.006 &    34.3 & Heintz 1974                &      \\
1973.84\phn\phn &       342.48       & 8.300          & \phn3 &  $-$0.3 &  $-$0.026 &    24.2 & Heintz 1974                &      \\
1974.809\phn    &       342.5\phn    & 8.41\phn       & \phn1 &  $-$0.0 & \phs0.037 &    14.0 & van Albada-van Dien 1983   & E    \\
1975.862\phn    &       342.25       & 8.438          & \phn1 & \phs0.0 & \phs0.016 &    14.0 & Josties et al.\ 1978       & E,G  \\
1975.892\phn    &       341.97       & 8.463          & \phn1 &  $-$0.2 & \phs0.039 &    14.0 & Josties et al.\ 1978       & E    \\
1975.927\phn    &       342.01       & 8.456          & \phn1 &  $-$0.2 & \phs0.031 &    14.0 & Josties et al.\ 1978       & E,G  \\
1976.047\phn    &       342.22       & 8.460          & \phn1 & \phs0.1 & \phs0.029 &    14.0 & Josties et al.\ 1978       & E,G  \\
1977.919\phn    &       340.3\phn    & 9.71\phn       & \phn1 &  $-$1.4 & \phs1.198 & \phn0.0 & Holden 1978                & A,E  \\
1982.661\phn    &       339.9\phn    & 8.97\phn       & \phn2 &  $-$0.5 & \phs0.284 & \phn9.7 & Argyle 1983                & E    \\
1988.101\phn    &       340.0\phn    & 8.10\phn       & \phn1 & \phs1.1 &  $-$0.725 & \phn7.2 & Popovic 1989               & E    \\
1988.23\phn\phn &       341.2\phn    & 8.93\phn       & \phn4 & \phs2.3 & \phs0.103 & \phn3.7 & Sturdy 1992                &      \\
1994.128\phn    &       337.5\phn    & 8.92\phn       & \phn1 & \phs0.1 & \phs0.022 &    20.0 & Abad \& Della Prugna 1995  & E,H  \\
1995.024\phn    &       336.82       & 8.89\phn       & \phn5 &  $-$0.4 &  $-$0.011 &    44.7 & Abad et al.\ 1998          & E    \\
2006.922\phn    &       333.72       & 8.781          & \phn1 &  $-$0.4 & \phs0.039 &    18.4 & Heinze et al.\ 2010        & E    \\
2009.036\phn    &       332.3\phn    & 8.53\phn       & \phn1 &  $-$1.3 &  $-$0.142 &    18.4 & Mason et al.\ 2011a        &      \\
2010.720\phn    &       332.8\phn    & 8.68\phn       & \phn2 &  $-$0.3 & \phs0.074 &    28.3 & Mason et al.\ 2011b        &      \\
2011.883\phn    &       332.23       & 8.05\phn       & \phn1 &  $-$0.6 &  $-$0.506 & \phn5.0 & Fay 2013                   & E,I  \\
2011.9903       &       330.5\phn    & 8.16\phn       & \phn1 &  $-$2.3 &  $-$0.391 & \phn5.0 & Micello 2012               & E,I  \\
2016.129\phn    &       330.41       & 8.332          & \phn1 &  $-$1.2 &  $-$0.003 &    20.0 & Locatelli 2017             & E    \\
\enddata
\tablenotetext{~}{A : Measure given zero weight in final orbit solution due to excessive residuals.}
\tablenotetext{~}{B : Measure by J.\ Gledhill cited by Lewis.}
\tablenotetext{~}{C : Measure uncertain or estimated by observer.}
\tablenotetext{~}{D : Number of nights varies 50\% or more between angle and separation measures. In this case, N = 
                  $\frac{N_{\theta}}{N_{\rho}}$, rounding down.}
\tablenotetext{~}{E : Original data published at the Besselian Epoch converted to the Julian Epoch as described in the
                  text.}
\tablenotetext{~}{F : Identification error in publication corrected.}
\tablenotetext{~}{G : Mean of multiple measures on the same photographic plate.}
\tablenotetext{~}{H : Quadrant flipped 180$^{\circ}$ from published value.}
\tablenotetext{~}{I : Measure given reduced weight in final orbit solution due to large residuals.}
\end{deluxetable}

\pagestyle{empty}
\def\msun{\hbox{${\cal{M}}_{\odot}$}}
\def\massA{\hbox{${\cal{M}}_A$}}
\def\massB{\hbox{${\cal{M}}_B$}}
\def\massAB{\hbox{${\cal{M}}_{A+B}$}}
\begin{deluxetable}{lc|cccc}
\rotate
\pagestyle{empty}
\tablenum{4}
\tablewidth{0pt}
\tablecaption{Orbital Elements of 40 Eri BC}
\tablehead{
\colhead{Element} & 
\colhead{New Orbit} &
\colhead{Heintz} &
\colhead{Wielen} &
\colhead{van den Bos} &
\colhead{Gore} \\
\colhead{~} & 
\colhead{~} &
\colhead{(1974)} &
\colhead{(1962)} &
\colhead{(1926)} &
\colhead{(1886)} \\
}
\startdata
\small Period; P (yrs)                                    &       \phn230.09\phn\phn$\pm$0.68\phn\phn       &       \phn252.1\phn\phn &       \phn251.988\phn$\pm$5.824\phn &       \phn247.92\phn\phn$\pm$9.7\phn\phn        &       \phn139.0\phn\phn \\
\small Semi-major axis; a ($''$)                          & \phn\phn\phn6.931\phn$\pm$0.050\phn             & \phn\phn\phn6.943       & \phn\phn\phn7.0453$\pm$0.0925       & \phn\phn\phn6.8945\phs\phn\phn\phn\phn          & \phn\phn\phn5.99\phn    \\
\small Inclination; i ($^{\circ}$)                        &       \phn107.53\phn\phn$\pm$0.29\phn\phn       &       \phn108.9\phn\phn &       \phn108.540\phn$\pm$0.375\phn &       \phn 71.55\phn\phn\phs\phn\phn\phn\phn    &    \phn\phn76.3\phn\phn \\
\small Longitude of Node; $\Omega$ ($^{\circ}$)           &       \phn151.44\phn\phn$\pm$0.12\phn\phn       &       \phn150.9\phn\phn &       \phn150.958\phn$\pm$0.426\phn &       \phn150.96\phn\phn\phs\phn\phn\phn\phn    &       \phn146.3\phn\phn \\
\small Epoch (2000) of Periastron; T$_{o}$ (yrs)          &          1847.6\phn\phn\phn$\pm$1.1\phn\phn\phn &          1849.6\phn\phn &          1848.872\phn$\pm$0.876\phn &          1848.93\phn\phn$\pm$0.93\phn           &          1863.88\phn    \\
\small Eccentricity; e                                    & \phn\phn\phn0.4300$\pm$0.0027                   & \phn\phn\phn0.410       & \phn\phn\phn0.4147$\pm$0.0100       & \phn\phn\phn0.4024$\pm$0.020                    & \phn\phn\phn0.136       \\
\small Longitude of Periastron; $\omega$ ($^{\circ}$)     &       \phn318.2\phn\phn\phn$\pm$1.1\phn\phn\phn &       \phn327.8\phn\phn &       \phn326.497\phn$\pm$1.765\phn &       \phn326.96\phn\phn\phs\phn\phn\phn\phn    &       \phn354.4\phn\phn \\
                                                          &                                                 &                         &                                     &                                                 &                         \\
\small Parallax (mas, van Leeuwen 2007)                   &       \phn200.62\phn\phn$\pm$0.23\phn\phn       &                         &                                     &                                                 &                         \\
\small Fractional Mass ($f = \frac{C}{B+C}$, Heintz 1974) & \phn\phn\phn0.262\phn$\pm$0.01\phn\phn          &                         &                                     &                                                 &                         \\
                                                          &                                                 &                         &                                     &                                                 &                         \\
\small \phn\phn White Dwarf Mass (\msun)                  & \phn\phn\phn0.575\phn$\pm$0.018\phn             & 0.43$\pm$0.02           & $\Sigma\cal{M}$=0.678$\pm$0.055     & \phn\phn\phn0.44\phn\phn$\pm$0.11\phn           & $\Sigma\cal{M}$=1.003   \\
\small \phn\phn Red Dwarf Mass (\msun)                    & \phn\phn\phn0.2041$\pm$0.0064                   & 0.16$\pm$0.01           &                                     & \phn\phn\phn0.20\phn\phn$\pm$0.05\phn           &                         \\
\enddata
\end{deluxetable}


\begin{deluxetable}{lcc}
\tablenum{5}
\tablewidth{0pt}
\tablecaption{Ephemerides of 40 Eri BC}
\tablehead{
\colhead{Epoch} &
\colhead{$\theta$} &
\colhead{$\rho$} \\
\colhead{~} &
\colhead{(deg)} &
\colhead{(arcsec)} 
}
\startdata
2018.0 & 331.0 & 8.219 \\
2019.0 & 330.7 & 8.151 \\
2020.0 & 330.4 & 8.081 \\
2021.0 & 330.0 & 8.007 \\
2022.0 & 329.7 & 7.929 \\
2023.0 & 329.4 & 7.847 \\
2024.0 & 329.1 & 7.762 \\
2025.0 & 328.7 & 7.674 \\
2026.0 & 328.4 & 7.581 \\
2027.0 & 328.0 & 7.484 \\
\enddata
\end{deluxetable}

\end{document}